\documentclass[aip,nofootinbib]{revtex4} 
\usepackage{amsmath}
\DeclareSymbolFont{AMSb}{U}{msb}{m}{n}
\DeclareMathSymbol{\R}{\mathalpha}{AMSb}{"52}

\begin{document}
\title{Upper limit on the critical strength of central potentials in relativistic quantum mechanics}
\author{Fabian \surname{Brau}}
\email[E-mail: ]{fabian.brau@umh.ac.be}
\affiliation{Groupe de Physique Nucl\'eaire Th\'eorique, Acad\'emie Universitaire Wallonie-Bruxelles, Universit\'e de Mons-Hainaut, B-7000 Mons, Belgique}

\begin{abstract}
In the context of relativistic quantum mechanics, where the Schr\"odinger equation is replaced by the spinless Salpeter equation, we show how to construct a large class of upper limits on the critical value, $g_{\rm{c}}^{(\ell)}$, of the coupling constant, $g$, of the central potential, $V(r)=-g v(r)$. This critical value is the value of $g$ for which a first $\ell$-wave bound state appears.
\end{abstract}


\maketitle 

\section{Introduction}
\label{sec1}
A covariant description of bound states of two particles is achieved with the Bethe-Salpeter equation \cite{salp51}. This equation reduces to the spinless Salpeter equation \cite{grei94} when the following approximations are performed:  
\begin{itemize}
\item elimination of any dependences on timelike variables (which leads to the Salpeter equation \cite{salp52}).
\item any references to the spin degrees of freedom of particles are neglected as well as negative energy solutions.
\end{itemize}
 
The spinless Salpeter equation takes the form ($\hbar=c=1$)
\begin{equation}
\label{eq1}
\left[\sqrt{{\bf p}^2+m^2}+V({\bf r})\right]\, \Psi({\bf r})=M\, \Psi({\bf r}),
\end{equation}
where $m$ is the mass of the particle and $M$ is the mass of the eigenstate ($M=m+E$, $E$ is the binding energy). We restrict our attention to interactions which are introduced in the free equation through the substitution $M \rightarrow M-V({\bf r})$, where $V({\bf r})$ is the time component of a relativistic four-vector. The interaction could also, in principle, be introduced through the substitution ${\bf p} \rightarrow {\bf p}-{\bf A}({\bf r})$, where ${\bf A}({\bf r})$ is the spatial component of a relativistic four-vector. However we do not consider this kind of potentials since the derivation of the spinless Salpeter equation from the Bethe-Salpeter equation leads to ${\bf A}({\bf r})=0$. Equation (\ref{eq1}) is generally used when kinetic relativistic effects cannot be neglected and when the particles under consideration are bosons or when the spin of the particles is neglected or is only taken into account via spin-dependent interactions. Despite its apparent complexity, this equation is often preferred to the Klein-Gordon equation. Equation (\ref{eq1}) appears, for example, in mesons and baryons spectroscopy in the context of potential models (see for example \cite{go85,se97,br98,gl98,br02}) [For a review of several aspects of the ``semirelativistic" description of bound states with the spinless Salpeter equation see: W. Lucha and F. F. Sch\"oberl, Int. J. Mod. Phys. A \textbf{14}, 2309 (1999) and references therein.] 

Due to the pseudo-differential nature of the kinetic energy operator, few exact results are known about this equation. Most of these results have been obtained for a Coulomb potential (for example, upper and lower bounds on energy levels) \cite{he77,ca84,ha84,ma89,ra94}. Recently, upper and lower limits on energy levels have also been obtained for some other particular interactions \cite{ha01a,ha01b,ha02,ha03}.

Conversely to the Schr\"odinger equation, for which a fairly large number of results giving both upper and lower limits on the number of bound states can be found in the literature (see for example \cite{ba52,sc61,ca65a,ca65b,ch68,gl76,ma77,br03a,br03b,br03c}), only two results are known for the spinless Salpeter equation \cite{daub83,br03d}. The first result, obtained in Ref. \cite{daub83}, is an upper bound on the total number of bound states yielding a lower limit on the critical value, $g_{\rm{c}}^{(0)}$, of the coupling constant (strength), $g$, for which a first S-wave ($\ell=0$) bound state appears ($\ell$ being obviously the angular momentum) in the potential $V(r)=-g v(r)$. The second results, obtained in Ref. \cite{br03d}, is an upper limit on the number of $\ell$-wave bound states which yields a lower limit on the critical value, $g_{\rm{c}}^{(\ell)}$ for which a first $\ell$-wave bound state appears. 

In this article, we obtain accurate upper limits on the critical strength $g_{\rm{c}}^{(\ell)}$ applicable to attractive (purely negative) central potentials which are less singular than $-r^{-1}$ at the origin. This limitation has a deep reason. Indeed, it is known that for the spinless Salpeter equation, a potential which behaves like $-r^{-1}$ at the origin is characterized by a maximal value of the coupling constant above which the spectrum is no longer bounded from below. This particularity has been studied in detail for the Coulomb potential (see for example \cite{he77}). The $-r^{-1}$ singularity is a critical singularity for the spinless Salpeter equation just as the $-r^{-2}$ singularity is a critical singularity for the Schr\"odinger equation. So in this article we discard this class of potentials which should be treated separately. Moreover we suppose that the central potential $V(r)$ is piecewise continuous for $r \in\ ]0,\infty[$. The upper limits on $g_{\rm{c}}^{(\ell)}$ we obtain in sec.~\ref{sec2} are compared with the exact critical value obtained numerically for some test potentials. These comparisons indicate that the new upper limits are very restrictive. Some conclusions are presented in sec.~\ref{sec3}.

\section{Upper limit on the critical strength}
\label{sec2}

The idea used to derived the upper limit on $g_{\rm{c}}^{(\ell)}$ is to transform the standard eigenvalue problem obtained with the time independent spinless Salpeter equation (\ref{eq1}), and where the eigenvalues are the eigenenergies, into an eigenvalue problem where the eigenvalues are the critical coupling constants. These critical values of the strength of the potential correspond to the occurrence of an eigenstate with a vanishing binding energy. We thus consider the zero binding energy spinless Salpeter equation that we need to write as an integral equation. This has be done in Ref. \cite{br03d} but since we need some modifications in the development, we recall the main line here. 

We have to calculate the Green function of the kinetic energy operator. Similar calculations have also already been performed previously \cite{nick84,brau98}. In contrast to results found in Ref.~\cite{brau98}, we need here to calculate the Green function of the following operator
\begin{equation}
\label{eq2}
T\left({\bf p}^2\right)=\sqrt{{\bf p}^2+m^2}-m.
\end{equation}
This is done by performing the integral
\begin{equation}
\label{eq3}
G(m,\Delta)=\frac{1}{(2\pi)^3}\int d{\bf p}\, \frac{\exp(-i\, {\bf p}\cdot {\bf \Delta})}
{\sqrt{p^2+m^2}-m},
\end{equation}
where ${\bf \Delta}={\bf r}-{\bf r}'$ and $\Delta=|{\bf \Delta}|$.
We find that
\begin{subequations}
\label{eq4}
\begin{equation}
\label{eq4a}
G(m,\Delta)=\frac{m}{4\pi \Delta}\left[1+\frac{2}{\pi}F(m\Delta)\right]\equiv 
\frac{m}{4\pi \Delta}\, H(m\Delta),
\end{equation}
with
\begin{equation}
\label{eq4b}
F(y)=\int_y^{\infty} \frac{dz}{z}\, K_1(z)+\frac{\pi}{2} 
    =K_1(y)+\frac{\pi}{2}-\int_y^{\infty} dz\, K_0(z),
\end{equation}
\end{subequations}
and where $K_{\nu}(y)$ is a modified Bessel function (see for example \cite[p. 374]{abra70}).
The zero binding energy spinless Salpeter equation takes thus the form of the following integral equation
\begin{equation}
\label{eq5}
\Psi({\bf r})=-\int d{\bf r}'\, G(m,\Delta)\, V({\bf r}')\, \Psi({\bf r}'),
\end{equation}
with $G(m,\Delta)$ given by (\ref{eq4}). We now restrict our attention to central potentials $ V({\bf r})= V(r)$, with $r=|{\bf r}|$.

Integration over angular variables reduces the integral equation (\ref{eq5}) to the following one-dimensional integral equation
\begin{subequations}
\label{eq6}
\begin{equation}
\label{eq6a}
u_{\ell}(r)=-\int_0^{\infty} dr'\, G_{\ell}(m,r,r')\, V(r')\, u_{\ell}(r'),
\end{equation}
with
\begin{equation}
\label{eq6b}
G_{\ell}(m,r,r')=\frac{mrr'}{2}\int_0^{\pi}d\theta' \, \sin \theta' \, \frac{H(m\Delta)}{\Delta}\, P_{\ell}(\cos \theta'),
\end{equation}
\end{subequations}
where $u_{\ell}(r)$ is the radial wave function, 
$\Psi({\bf r})=(u_{\ell}(r)/r)Y_{\ell m}({\bf \hat{r}})$ and where $H(x)$ is defined by (\ref{eq4a}).

An important technical difficulty, to obtain a symmetrical kernel, appears if the potential possesses some change of sign  (see relation (\ref{eq7}) below). This is overcome when one searches for necessary conditions, or upper bound on the number of bound states, by replacing the potential by its negative part $V(r)\rightarrow V^-(r)=\max(0,-V(r))$. Indeed, the potential $V^-(r)$ is more attractive than $V(r)$ and thus a necessary condition for existence of bound states in $V^-(r)$ is certainly a valid necessary condition for $V(r)$. This procedure can no longer be used to obtain sufficient conditions. For this reason we consider potentials that are nowhere positive, $V(r)=-g v(r)$, with $v(r)\geq 0$.

The integral equation (\ref{eq6}) can be written with a symmetrical kernel provided we introduce a new wave function
\begin{equation}
\label{eq7}
\phi_{\ell}(r)=|V(r)|^{1/2}\, u_{\ell}(r).
\end{equation}
This change of function leads to the following integral equation
\begin{subequations}
\label{eq8}
\begin{equation}
\label{eq8a}
\phi_{\ell}(r)=g\int_0^{\infty}dr'\, K_{\ell}(m,r,r')\, \phi_{\ell}(r'),
\end{equation}
with
\begin{equation}
\label{eq8b}
K_{\ell}(m,r,r')=v(r)^{1/2}\, G_{\ell}(m,r,r')\,v(r')^{1/2}.
\end{equation}
\end{subequations}
The relation (\ref{eq8}) is thus an eigenvalue problem and, for each value of $\ell$, the smallest characteristic number is just the critical value $g_{\rm{c}}^{(\ell)}$. The other characteristic numbers correspond to the critical values of the strength of the potential for which a second, a third, ..., $\ell$-wave bound state appears. The kernel (\ref{eq8b}) acting on the Hilbert space $L^2(\R)$ is an Hilbert-Schmidt operator for potentials which decrease faster than $r^{-1}$ at infinity. Thus this kernel satisfies the inequality
\begin{equation}
\int_0^{\infty} \int_0^{\infty} dx\, dy\, K_{\ell}(x,y)K_{\ell}(x,y) < \infty.
\end{equation}
Consequently the eigenvalue problem (\ref{eq8}) always possesses at least one characteristic number \cite[pp. 102-106]{tri65} (in general, this problem has an infinity of characteristic numbers).

Now we use the theorem (see for example \cite[pp. 118-119]{tri65}) which states that, for a symmetric Hilbert-Schmidt kernel, we have the variational principle 
\begin{equation}
\label{eq10}
\max_{\varphi}\left|\int_0^{\infty}dx\, dy\, K_{\ell}(x,y)\, \varphi(x) \varphi(y)\right| =\frac{1}{|g_1|},
\end{equation}
for $\varphi(r)$ satisfying
\begin{equation}
\label{eq9}
\int_0^{\infty}dr\, \varphi(r)^2=1.
\end{equation}
The maximal value is reached for $\varphi(x)=\varphi_1(x)$, where $\varphi_1(x)$ is the eigenfunction associated to the smallest eigenvalue $g_1$. Consequently, for an arbitrary normalized function, $f(x)$, we obtain the following upper limit on $g_1$
\begin{equation}
\label{eq11}
|g_1|\leq\left|\int_0^{\infty}dx\, dy\, K_{\ell}(x,y)\, f(x) f(y)\right|^{-1}.
\end{equation}

For the clarity of the discussion we now consider in two separate sections the ultrarelativistic regime ($m=0$) and the relativistic regime ($m>0$).

\subsection{Ultrarelativistic regime $m=0$}
\label{sec2.1}

In this section, we derive an (among others) upper limit on the critical value, $g_{\rm{c}}^{(\ell)}$, of the coupling constant, $g$, of the potential, $V(r)=-g v(r)$, for which a first $\ell$-wave bound state appears in the ultrarelativistic regime ($m=0$). In this limit, the kernel takes a simple form since $m K_1(m y)=1/y$ when $m$ goes to zero. This implies that
\begin{equation}
\label{eq12}
\lim_{m\rightarrow 0} m H(m\Delta)=\frac{2}{\pi \Delta}.
\end{equation}
The function $G_{\ell}(0,r,r')$ takes then the form
\begin{equation}
\label{eq13}
G_{\ell}(0,r,r')=\frac{rr'}{\pi}\int_0^{\pi}d\theta' \, \frac{\sin \theta'}{\Delta^2}\, P_{\ell}(\cos \theta').
\end{equation}
A simple change of variable leads to \cite[p. 335]{abra70}
\begin{eqnarray}
\label{eq14}
G_{\ell}(0,r,r')&=&\frac{1}{2\pi}\int_{-1}^{1}dy \, \frac{P_{\ell}(y)}{(r^2+r'^2)/(2r r')-y}\nonumber \\
&=& \frac{1}{\pi} Q_{\ell}\left(\frac{r^2+r'^2}{2r r'}\right),
\end{eqnarray}
where the function $Q_{\ell}(x)$ is a Legendre function of the second kind. The function 
$G_{\ell}(0,r,r')$ can thus be evaluated explicitly for each value of the angular momentum $\ell$. We have for example
\begin{equation}
\label{eq15}
G_{0}(0,r,r')=\frac{1}{\pi}\ln\left|\frac{r+r'}{r-r'}\right|,
\end{equation}
and
\begin{equation}
\label{eq16}
G_{1}(0,r,r')=\frac{1}{\pi}\left[\frac{r^2+r'^2}{2 r r'}\ln\left|\frac{r+r'}{r-r'}\right|-1\right].
\end{equation}
Since the function $G_{\ell}(0,r,r')$ is given by the relation (\ref{eq14}), it follows that the kernel $K_{\ell}(0,r,r')$, see (\ref{eq8b}), is known for each value of $\ell$. Now, we just need to choose a suitable normalized function $f(r)$ to apply the variational principle reported above.

For simplicity we restrict the rest of the following discussion to $\ell=0$ but extensions to non vanishing values of the angular momentum is obvious, one just need to computed the corresponding expression of the function $G_{\ell}(0,r,r')$.

The function $f(r)$ should be as close as possible to the zero binding energy wave function but also should be general and simple enough to obtain a neat formula. We simply choose 
\begin{equation}
\label{eq17}
f(r)=A \left[r^{p-1}v(r)^p\right]^{1/2}, \quad p>0,
\end{equation}
where $A$ is the normalization factor. The relations (\ref{eq8b}), (\ref{eq11}), (\ref{eq15}) and (\ref{eq17}) lead to the following upper limit on $g_{\rm{c}}^{(0)}$
\begin{equation}
\label{eq18}
g_{\rm{c}}^{(0)}\leq \frac{\alpha \pi\int_0^{\infty}dx\, F_1(2p-1;x)}{2\int_0^{\infty}dx\, F_1(p;x) \int_0^x dy\, F_1(p;y) \ln\left(\frac{x+y}{x-y}\right)}\equiv g_{\rm{up},1}^{m=0},
\end{equation}
where $F_1(q;x)=x^{(q-1)/2} v(x)^{(q+1)/2}$ and where we have introduced the parameter $\alpha$ which takes the value 1 respectively 2 for one respectively two (identical) particle problems. The most stringent upper limit is obviously obtained by minimizing the right hand side of (\ref{eq18}) with respect to all positive values of $p$.

A simpler, but less stringent, version of this upper limit can be obtained with the help of the following minorization
\begin{equation}
\label{eq19}
\ln\left(\frac{x+y}{x-y}\right)\geq \frac{2y}{x}.
\end{equation}
\begin{equation}
\label{eq20}
g_{\rm{c}}^{(0)}\leq \frac{\alpha \pi\int_0^{\infty}dx\, F_1(2p-1;x)}{4\int_0^{\infty}dx\, 
x^{-1}\, F_1(p;x) \int_0^x dy\, y\, F_1(p;y) }\equiv g_{\rm{up},2}^{m=0}.
\end{equation}

The accuracy of these upper limits can be tested with some typical potentials. The comparison between the exact results (obtained by solving numerically the spinless Salpeter equation) and the upper limits (\ref{eq18}) and (\ref{eq20}) is reported in Table \ref{tab1}. We have also added two lower limits on $g_{\rm{c}}^{(0)}$ obtained with the upper limits on the number of bound states derived in Refs. \cite{daub83,br03d}. Note that for these tests, we choose a two identical particles problem, $\alpha=2$.

The results reported in Table \ref{tab1} indicate clearly that the accuracy of the upper limit (\ref{eq18}) is quite remarkable. The upper limit (\ref{eq20}) is obviously less stringent but could prove to be useful to obtain explicit formulas. The typical value of $p$ which optimize these upper limits varies between 2 and 3. We do not consider other choices for $f(r)$ (see (\ref{eq17})) here since the relation (\ref{eq18}) is already very accurate.

As an additional indication that the upper limits obtained with the method proposed in this work are quite accurate, we report in Table \ref{tab2} a comparison between the exact value of the critical strength $g_{\rm{c}}^{(1)}$ ($\ell=1$) and the corresponding upper limit obtained with the relations (\ref{eq11}), (\ref{eq16}) and (\ref{eq17}) and noted 
$g_{\rm{up}}^{m=0,\ell=1}$ in this Table.

\subsection{Relativistic regime $m>0$}
\label{sec2.2}

To obtain an upper limit on $g_{\rm{c}}^{(\ell)}$ for a non vanishing mass $m$, we need to calculate the expression of the function $G_{\ell}(m,r,r')$. To this end we note that
\begin{equation}
\label{eq21}
K_1(y)\leq F(y)\leq K_1(y)+\frac{\pi}{2}.
\end{equation}
>From the relation (\ref{eq11}), it is obvious that a minorization of the kernel $K_{\ell}(m,r,r')$ is enough to obtain the upper limit. However, the minorization (\ref{eq21}) of the function $F(y)$ is too crude to obtain good results. Instead we use
\begin{equation}
\label{eq21b}
F(y)\geq K_1(y)+\frac{\pi}{2}-\frac{\pi}{2} \exp(-y).
\end{equation}
This minorization (\ref{eq21b}) is proved in Appendix \ref{app1}. From the definition of $G_{\ell}(m,r,r')$ (\ref{eq6b}) and the inequality (\ref{eq21b}) we obtain
\begin{equation}
\label{eq22}
G_{\ell}(m,r,r')\geq \frac{1}{\pi}{\cal G}_{\ell}(m,r,r')+{\cal S}_{\ell}(m,r,r')-\frac{1}{2}{\cal T}_{\ell}(m,r,r'),
\end{equation}
where
\begin{equation}
\label{eq23}
{\cal G}_{\ell}(m,r,r')=m\int_{|r-r'|}^{r+r'}dy\, K_1(my)\, P_{\ell}\left(\frac{r^2+r'^2-y^2}{2rr'}\right),
\end{equation}
\begin{eqnarray}
\label{eq24}
{\cal S}_{\ell}(m,r,r')&=&m\int_{|r-r'|}^{r+r'}dy\, P_{\ell}\left(\frac{r^2+r'^2-y^2}{2rr'}\right), \nonumber \\ 
&=&\frac{2m}{2\ell+1}\, r_<^{\ell+1}\, r_>^{-\ell},
\end{eqnarray}
with $r_<=\min[r,r']$ and $r_>=\max[r,r']$ and \cite{magn66}
\begin{eqnarray}
\label{eq24b}
{\cal T}_{\ell}(m,r,r')&=&mr r'\int_{-1}^{1}dy\, \frac{\exp\left(-m\sqrt{r^2+r'^2-2r r' y}\right)}{\sqrt{r^2+r'^2-2r r' y}}\, P_{\ell}(y), \nonumber \\
&=& \sqrt{\frac{2}{\pi}} m^2 r r' \int_{-1}^{1}dy\, \frac{K_{1/2}\left(-m\sqrt{r^2+r'^2-2r r' y}\right)}{[m^2(r^2+r'^2-2r r' y)]^{1/4}}\, P_{\ell}(y), \nonumber \\
&=& 2m\sqrt{r r'}\,  K_{\ell+\frac{1}{2}}(m r_>)\, I_{\ell+\frac{1}{2}}(m r_<),
\end{eqnarray}
where $I_{\nu}(x)$ is a modified Bessel function (see for example \cite[p. 374]{abra70}). The kernel ${\cal S}_{\ell}(m,r,r')$ is actually the Green function of the nonrelativistic kinetic energy operator and takes a simple form while the kernel ${\cal G}_{\ell}(m,r,r')$ can be calculated analytically for each value of $\ell$ \cite{nick84,brau98}. We find for example
\begin{eqnarray}
\label{eq25}
{\cal G}_0(m,r,r')&=&K_0(m|r-r'|)-K_0(m(r+r')), \\
\label{eq26}
 {\cal G}_1(m,r,r')&=& K_0(m|r-r'|)+K_0(m(r+r')) \nonumber \\ &+& \frac{1}{m r r'} \left[(r+r')\, K_1(m(r+r'))-|r-r'|\, K_1(m|r-r'|)\right].
\end{eqnarray}

Now, we just need to choose a suitable normalized function $f(r)$ to apply the variational principle reported above (see (\ref{eq11})). We take the following expression for $f(r)$
\begin{equation}
\label{eq26b}
f(r)=A \left[r^{2p-1}v(r)^p\right]^{1/2}, \quad p>0.
\end{equation}

For simplicity we again restrict the rest of the following discussion to $\ell=0$ but extensions to non vanishing values of the angular momentum is obvious, one just need to computed the corresponding expression of the function $G_{\ell}(m,r,r')$.

The relations (\ref{eq8b}), (\ref{eq11}), (\ref{eq17}) and (\ref{eq25}) lead to the following upper limit on $g_{\rm{c}}^{(0)}$
\begin{subequations}
\label{eq27}
\begin{equation}
\label{eq27a}
g_{\rm{c}}^{(0)}\leq \frac{\alpha \int_0^{\infty}dx\, \sqrt{x} F_2(2p-1;x)}{2\int_0^{\infty}dx\, F_2(p;x) \int_0^x dy\, F_2(p,y) T(x,y)}\equiv g_{\rm{up}}^{m>0},
\end{equation}
with 
\begin{equation}
\label{eq27b}
T(x,y)=\frac{1}{\pi}[K_0(\beta(x-y))-K_0(\beta(x+y))]+2\beta y+\frac{1}{2}[\exp(-\beta(x+y))-\exp(-\beta(x-y))],
\end{equation}
\end{subequations} 
where $F_2(q,x)=x^{q-1/2} v(x)^{(q+1)/2}$ and where $\beta= mR$, $R$ being the scale of length which appears in the potential ($v(r)=v(Rx)$). Again, we have introduced in (\ref{eq27a}) the parameter $\alpha$ which takes the value 1 respectively 2 for one respectively two (identical) particle problems. 

The accuracy of this upper limit can be tested with some typical potentials. The comparison between the exact results and the upper limit (\ref{eq27}) is reported in Table \ref{tab3}. Note that for these tests, we also choose a two identical particles problem, $\alpha=2$.

The results reported in Table \ref{tab3} indicate clearly that the accuracy of the upper limit (\ref{eq27}) is quite good. But for small value of $\beta$ the upper limit is however less restrictive. Thus for small value of $\beta$ it is preferable to use an intermediate form for $f(r)$. We then propose in general to use
\begin{equation}
\label{eq28}
f(r)=A \left[r^{ap-1}v(r)^p\right]^{1/2}, \quad p>0,
\end{equation}
with $1\leq a\leq 2$. This last expression for $f(r)$ improves significantly the restriction on the possible values of $g_{\rm{c}}^{(0)}$. Indeed, for $\beta=0.1$, $a=1.18$ and for the exponential potential, the upper limit is then equal to $4.812$ instead of $5.390$. But this additional flexibility is only significant for small value of $\beta$, indeed for $\beta=0.5$ the best upper limit is found to be equal to $2.521$ (for the exponential potential and $a=1.69$) instead of $2.547$. 

However, even with the choice (\ref{eq28}) for $f(r)$, the upper limit (\ref{eq27}) still yields less restrictive results for small $\beta$ than those obtained for larger values of $\beta$ or those obtained with the upper limit (\ref{eq18}). This is easy to understand, since this is in the sector of small $\beta$ that the error introduced by the inequality (\ref{eq21b}) is the most important. Indeed, in the limit of $\beta$ going to zero, the upper limit (\ref{eq27}) coincides with the upper limit (\ref{eq18}) and for $\beta$ going to infinity, only the non relativistic kernel ${\cal S}_{\ell}(m,r,r')$ contributes. 

\section{Conclusions}
\label{sec3}

In this article we have shown how to construct upper limits on the critical value, $g_{\rm{c}}^{(\ell)}$, of the coupling constant, $g$, of a central potential, $V(r)=-g v(r)$. The method used to derive the upper limits is quite general and other (possibly more complicated) families of upper limits yielding (possibly) stronger restrictions on $g_{\rm{c}}^{(\ell)}$ could also be obtained. Indeed, the method is based on a variational principle for which a trial zero energy wave function is needed. There is no limitation on the accuracy of such a trial function, which imply that there is, in principle, no limitation on the accuracy of the upper limit on $g_{\rm{c}}^{(\ell)}$ derived with this procedure. However, this remark is only true for the ultrarelativistic regime, $m=0$, where the kernel of the integral equation has been calculated exactly. For $m>0$, a minorization of the kernel has been used yielding some errors in the restrictions on the possible values of the critical value $g_{\rm{c}}^{(\ell)}$ which cannot be compensated by a better choice of the trial zero energy wave function. In this article we have proposed in sec.~\ref{sec2} a compromise between accuracy and simplicity of the final formula. The accuracy of the upper limits on $g_{\rm{c}}^{(\ell)}$ was then tested with some typical potentials.

\begin{acknowledgments}
We thank the referee for his valuable comments on the manuscript. This work was supported by the National Funds for Scientific Research (FNRS), Belgium.
\end{acknowledgments}

\appendix
\section{Majorization of the primitive of $K_0(x)$}
\label{app1}

We choose the following integral representation for the modified Bessel function $K_0(x)$ \cite[p. 376]{abra70}
\begin{equation}
\label{eqa1}
K_0(x)=\int_0^{\infty}dt \exp(-x \cosh t).
\end{equation} 
We have
\begin{equation}
\label{eqa2}
\int_y^{\infty} dx\, K_0(x)=\int_0^{\infty}dt\, \frac{\exp(-y \cosh t)}{\cosh t}\leq \exp(-y) \int_0^{\infty}dt\, \frac{1}{\cosh t}= \frac{\pi}{2}\exp(-y).
\end{equation}

\clearpage

\begin{table}
\protect\caption{Comparison, for some typical potentials, between the exact critical values, $g_{\rm{c}}^{(0)}$, the upper limits $g_{\rm{up},1}^{m=0}$ (\protect\ref{eq18}), $g_{\rm{up},2}^{m=0}$ (\protect\ref{eq20}) and the lower limits obtained in Refs. \protect\cite{daub83,br03d}.}
\label{tab1}
\begin{tabular}{cccccc}
\hline
\hline
$v(x)$  & Ref. \cite{br03d} & Ref. \cite{daub83} & $g_{\rm{c}}^{(0)}$ & $g_{\rm{up},1}^{m=0}$ &$g_{\rm{up},2}^{m=0}$ \\
\hline
$\exp(-x)$        & 4.443 & 4.370 & 5.574 & 5.584 & 7.411 \\
$[\cosh(x)]^{-2}$ & 4.126 & 3.886 & 5.008 & 5.018 & 6.769 \\
$\exp(-x^2)$      & 4.513 & 4.169 & 5.426 & 5.442 & 7.399 \\
$x\exp(-x)$       & 3.696 & 3.349 & 4.360 & 4.364 & 5.964 \\
\hline
\hline
\end{tabular}
\end{table}

\begin{table}
\protect\caption{Comparison, for some typical potentials, between the exact critical values, $g_{\rm{c}}^{(1)}$ and the upper limit $g_{\rm{up}}^{m=0,\ell=1}$ obtained with the relations 
(\protect\ref{eq11}), (\protect\ref{eq16}) and (\protect\ref{eq17}).}
\label{tab2}
\begin{tabular}{ccc}
\hline
\hline
$v(x)$  & $g_{\rm{c}}^{(1)}$ & $g_{\rm{up}}^{m=0,\ell=1}$ \\
\hline
$\exp(-x)$        & 10.975 & 10.992 \\
$[\cosh(x)]^{-2}$ & 8.1174 & 8.1268 \\
$\exp(-x^2)$      & 10.200 & 10.231 \\
$x\exp(-x)$       & 9.5442 & 9.5636 \\
\hline
\hline
\end{tabular}
\end{table}

\begin{table}
\protect\caption{Comparison, for some typical potentials, between the exact critical values, $g_{\rm{c}}^{(0)}$ and the upper limit $g_{\rm{up}}^{m>0}$ (\protect\ref{eq27}).}
\label{tab3}
\begin{tabular}{ccccccc}
\hline
\hline
$v(x)$ & \multicolumn{2}{c}{$\exp(-x)$} & \multicolumn{2}{c}{$[\cosh(x)]^{-2}$} & \multicolumn{2}{c}{$\exp(-x^2)$} \\
\cline{1-7}
$\beta$  &$g_{\rm{c}}^{(0)}$  &$g_{\rm{up}}^{m>0}$ & $g_{\rm{c}}^{(0)}$ & $g_{\rm{up}}^{m>0}$ &$g_{\rm{c}}^{(0)}$& $g_{\rm{up}}^{m>0}$\\
\hline
0.1 &  4.694  & 5.390  & 4.461  & 4.994  & 4.927  & 5.363 \\
0.5 &  2.387  & 2.547  & 2.766  & 3.006  & 3.309  & 3.589 \\
1   &  1.361  & 1.407  & 1.742  & 1.843  & 2.198  & 2.352 \\
2   &  0.7133 & 0.7206 & 0.9598 & 0.9862 & 1.257  & 1.307 \\
3   &  0.4804 & 0.4817 & 0.6549 & 0.6642 & 0.8669 & 0.8880 \\
4   &  0.3607 & 0.3615 & 0.4956 & 0.4994 & 0.6589 & 0.6694 \\
5   &  0.2890 & 0.2893 & 0.3981 & 0.3999 & 0.5305 & 0.5364 \\
\hline
\hline
\end{tabular}
\end{table}

\end{document}